\documentclass[webconf, onecolumn]{svjour}

\usepackage{ifpdf}
\ifpdf
  \usepackage[pdftex]{graphicx,color}
\else
  \usepackage[dvips]{graphicx,color}
\fi

\usepackage{graphicx}
\usepackage[varg]{txfonts} 
\usepackage[latin1]{inputenc}
\DeclareGraphicsExtensions{.eps,.eps.gz,.ps,.fig,.jpg,.png}
\session-title{First conference on Adaptive Optics for Extremely Large Telescopes}
\begin{document}
\title{EAGLE multi-object AO concept study for the E-ELT}
\author{G. Rousset\inst{1}\fnmsep\thanks{\email{gerard.rousset@obspm.fr}} \and T. Fusco\inst{2} 
\and F. Assemat\inst{1,2} \and T. Morris\inst{3} \and E. Gendron\inst{1}  \and R. Myers\inst{3}
\and M. Brangier\inst{1}  
\and M. Cohen\inst{4}  \and N. Dipper\inst{3} \and C. Evans\inst{5} \and D. Gratadour\inst{1}  
\and P. Jagourel\inst{4}  \and P. Laporte\inst{4}  \and D. Le Mignant\inst{6} 
\and M. Puech\inst{4} \and C. Robert\inst{2}  \and H. Schnetler\inst{5}  \and W. Taylor\inst{5} 
\and F. Vidal\inst{1} 
\and J.-G. Cuby\inst{6}   \and M. Lehnert\inst{4} \and S. Morris\inst{3}  \and P. Parr-Burman\inst{5}
}
\institute{LESIA, Observatoire de Paris, Universit\'e Paris Diderot, 5 place J. Janssen, 92190 Meudon, France
\and ONERA, 29 Av. de la division Leclerc, 92322, Ch\^atillon, France 
\and Durham University, Department of Physics, South Road, Durham DH1 3LE, UK \and GEPI, Observatoire de Paris, Universit\'e Paris Diderot, 5 place J. Janssen, 92190 Meudon, France \and UK Astronomy Technology Centre, Blackford Hill, Edinburgh EH9 3HJ, UK 
\and LAM, Technop\^ole de Marseille-Etoile, 38 rue F. Joliot-Curie, 13388 Marseille cedex 13
}
\abstract{
EAGLE is the multi-object, spatially-resolved, near-IR spectrograph
instrument concept for the E-ELT, relying on a distributed Adaptive
Optics, so-called Multi Object Adaptive Optics. This paper presents
the results of a phase A study. Using $84$x$84$ actuator deformable
mirrors, the performed analysis demonstrates that $6$ laser guide
stars and up to $5$ natural guide stars of magnitude $R<17$, picked-up
in a $7.3'$ diameter patrol field of view, allow us to obtain an
overall performance in terms of Ensquared Energy of $35\%$ in a
$75$x$75 mas^2$ spaxel at $H$ band whatever the target direction in
the centred $5'$ science field for median seeing conditions. The
computed sky coverage at galactic latitudes $|b|\sim60$ is close to
$90\%$.  }
\maketitle
\section{Introduction}
\label{intro}
The European Extremely Large Telescope (E-ELT) is currently in phase B
at ESO. This phase will end mid 2010 with the release of the proposal
for the E-ELT construction. Meanwhile ESO has launched a number of
instrument conceptual studies (phases A)\cite{Kissler}. A high
priority instrument as derived from the science cases of the E-ELT, is
a near IR spectrograph with multi, deployable Integral Field Units
(IFUs), assisted by Adaptive Optics (AO). This type of instrument is
particularly required for the study of the evolution of galaxies
across cosmic times, addressing the key science areas on the physics
and evolution of high-redshift galaxies, the detection and
characterisation of first-light galaxies and the physics of galaxy
evolution from stellar archaeology\cite{Puech,Evans}. These are the
main scientific drivers of EAGLE (Elt Adaptive optics for GaLaxy
Evolution), the multi-object spatially-resolved near-IR spectrograph
concept for the E-ELT~\cite{Cuby}. EAGLE is a relatively simple
instrument relying on a distributed AO concept, so-called MOAO (Multi
Object AO).  The EAGLE consortium consists of six institutes in France
and in the UK. The EAGLE Phase A study started mid-2007 and ended in
October 2009. But prototyping and demonstration activities will
continue throughout 2010 and beyond. We present in this paper the
phase A concept study of the AO system of this instrument.

\section{Instrument requirements and interfaces}
\label{Inst}

EAGLE is motivated by the desire to obtain near-IR spectroscopy of
large numbers of objects across a wide field of view (FoV) of diameter
$>5'$, to build-up representative and unbiased samples of, for
example, hundreds of high-redshift galaxies (see
Ref.~\cite{Cuby}). The requirement is to implement 20 IFUs in parallel
and distributed in science FoV. It results from the trade-off between
complexity/cost and observing time.  The science case calls for
improved angular resolution ($75 mas$) with respect to the seeing but
not diffraction-limited performance. The main scientific drivers are
the improvement of the point source sensitivity and the ability to
spatially resolve structures at the $100 mas$ level. The performance
requirement has therefore been set as $30$ to $40\%$ Ensquared Energy
(EE) in a square spaxel of $75$x$75 mas^2$ in $H$ band ($1.6\mu
m$). The subfield dedicated to each IFU is set to $1.65''$x$1.65''$
and sampled at $37.5mas$. The wavelength coverage of the spectrographs
extends from $0.8$ to $2.5\mu m$.

The E-ELT baseline Design considers a 42-m telescope with a 5-mirror
concept, mirrors active and adaptive. EAGLE is planned for
installation at the Gravity Invariant Focal Station (GIFS) below the
Nasmyth B platform allowing us to reduce significantly the problems
associated with gravity induced flexure. The implementation
contemplates a large retractable M6 mirror bending the full $10'$
diameter telescope FoV down to the GIFS. As a result of this model the
instrument will need to take full control of the telescope, including
the wavefront sensing Natural Guide Stars for controlling the
telescope functions, such as co-phasing, telescope guiding and
tracking, field stabilisation and active optics.

\begin{figure}
\begin{center}
\resizebox{0.5\columnwidth}{!}{ \includegraphics{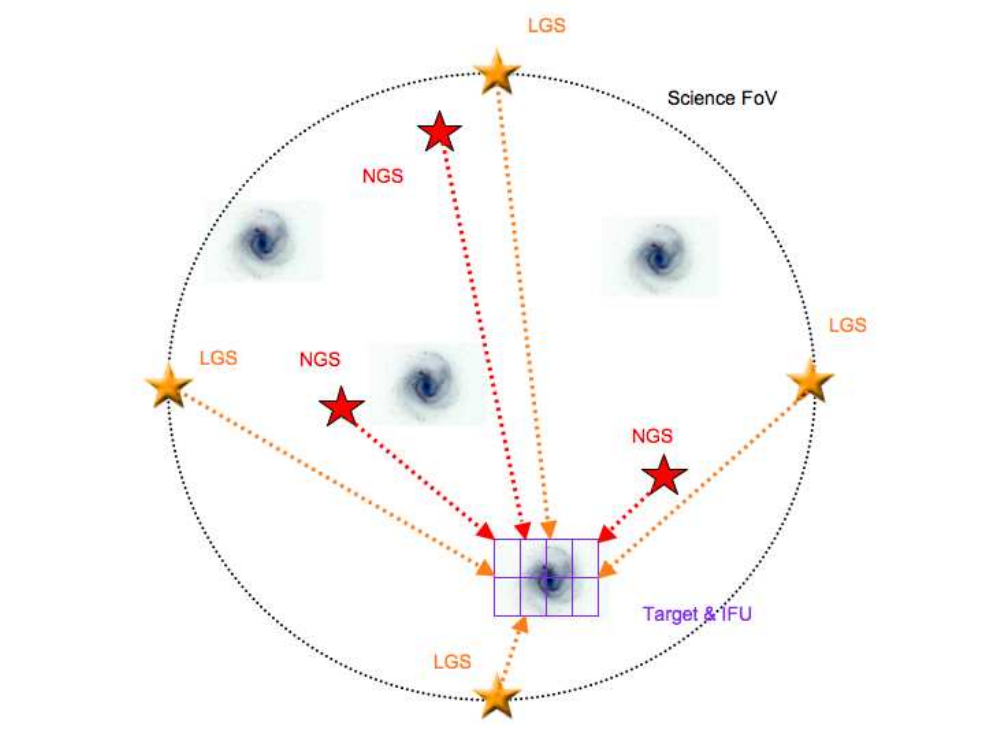} }
\caption{MOAO configuration for LGS, NGS and science FoV ; IFU on target including a dedicated DM controlled in open loop through tomography using all the GS measurements (the arrows)}
\label{moao}   
\end{center}
\end{figure}

The first performance driver for the EAGLE AO is this Ensquared Energy
(EE) requirement. The second one is the patrol FoV in the range of $5$
to $10'$ in diameter where the science targets are looking for and EE
shall be achieved.  Compensating for turbulence in such a wide FoV is
a real challenge in AO. Despite it simplicity, Ground Layer AO will
not achieve the EE requirement over the full patrol FoV considered for
EAGLE~\cite{Fusco}. Even Multi Conjugate AO (with a field
segmentation) is not affordable due to the substantial increase of the
number of deformable mirrors (DM) required to obtain the EE
specification~\cite{Fusco}. In fact, there is no need for a full
correction of the entire FoV but rather only in the specific
directions of the targets. Therefore, the Multi Object AO
(MOAO)~\cite{Assemat-F} is the only tractable concept allowing us to
achieve the EE performance as demonstrated in this paper. As presented
in Figure~\ref{moao}, the idea is to use one DM per direction of
interest (target) in order to only compensate the IFU subfield of size
$1.65''$x$1.65''$. For EAGLE, the science targets are too faint to
allow any wavefront (WF) measurement. The correction to be applied to
the science DMs is then computed by tomography from a set of guide
stars (GS) distributed in the whole patrol FoV, on which the turbulent
WFs are measured. These GS can be natural (NGS) or laser generated
(LGS). Hence the control of the science DMs is done in open loop.

The first optical element of EAGLE in the telescope focal plane is a
small $45\deg$ pick-off mirror to extract either the science target or
the NGS, sending the light to the re-imaging optical train towards the
IFU or the WF sensors (WFS), respectively. The science DMs are
therefore implemented in the optical trains feeding the IFUs. The DM
are conjugated to the telescope M4 mirror. Because of the large
footprints of the LGS, they are extracted just above the focal plane
with large pick-off mirrors, placed outside a preserved unvignetted
central science FoV of $5'$. The positions of the LGS are on an outer
ring as depicted in Fig.~\ref{moao}. The EAGLE AO system will take
advantage of E-ELT concept by controlling in closed loop the two
in-telescope adaptive mirrors: M4 (deformable) and M5 (field
stabilisation). It leads to a substantial reduction of the stroke
required for the science DMs.
 
\section{AO system analysis}
\label{analysis}

The relatively large spatial resolution element, when compared to that
of a diffraction limited instrument (approximately ten times larger),
has a huge influence on the design of AO and the error budget. Low
Order (LO) modes (such as Tip-Tilt and defocus for instance) have very
little impact on the overall performance of the system in terms of EE.
On the other hand, the only way to improve EE is to correct the High
Order (HO) spatial frequencies, requiring HO DMs and efficient
tomographic reconstruction. For the numerical simulation, we use a WF
power spectral density based tool~\cite{Neichel,Assemat08} including
multi-GS analysis, LGS and NGS WFSs, WF low order filtering for LGSs,
optimal tomography reconstruction, dedicated FoV direction DM
projection, point spread function (PSF) and EE computation.  In this
paper, the performance evaluation of EAGLE AO includes the tomographic
error linked to the science FoV and the number and position of GSs,
the propagated noise from the WFSs for both LGSs and NGSs, the fitting
and aliasing effects due to the selected number of subapertures and
actuators. The performance is EE, computed in $75$x$75 mas$ from a set
of PSFs at $H$ band, every $30''$ into the $5'$ diameter FoV.

A first analysis, presented elsewhere~\cite{Fusco}, lead to the
requirements to use 9 LGSs for tomography, one NGS to measure the very
low orders of the WF and high order science DMs. For cost and risk
reduction purposes, we explored other configurations reducing the
number of LGSs while taking advantage of the NGSs available in the FoV
of the instrument. We investigate here the simultaneous variability of
the following parameters:
\begin{itemize}
\item number of LGSs, $4$ and $6$ ; 
\item diameter of the FoV where NGSs could be picked-off, $5$, $7.3$ and $10'$ ; 
\item limiting magnitude of the NGSs for WF sensing, $R=15$, $17$ and $19$ ; 
\item maximum number of available NGS-dedicated WFSs in the instrument, $1$, $3$ and $5$.
\end{itemize}
We first assumed a seeing of $0.95''$ at $0.5\mu m$ in the line of
sight, an outer scale of $50m$, ten turbulence layers between $0$ and
$16.5 km$, a low noise WF measurement variance of $0.1 rad^2$ at
$0.5\mu m$ for the LGSs, the filtering of the Zernike radial orders 1
and 2 on the LGSs, a noise WF measurement variance of $1 rad^2$ at
$0.5\mu m$ for a NGS of magnitude $R\simeq11$ and a pupil sampling of
112 subapertures (for both LGS and NGS) and actuators in the diameter.
For the 4 LGS case, EE in the central $5'$ science FoV is only around
$30\%$ even considering 5 NGSs up to magnitude $19$ in a patrol field
of $10'$. These values do not provide enough margin to include in the
WF error (WFE) budget of the instrument all the other contributors not
simulated here and therefore to be compliant with the specification.
Another conclusion is that there is really a substantial gain in
performance when increasing the patrol FoV to the largest possible
value and increasing the number of available WFSs up to 5 because more
NGSs are available to complement the measurements on the LGSs. But
looking for NGSs up to $R=19$ does not bring any gain due to too low a
signal to noise ratio for such NGSs in the WF measurements.

In Figure~\ref{15-subfields}, we present the 6 LGS case considering a
true cosmologic field of interest, the XMM-LSS field as an example,
taking into account the sky coverage issue linked to the availability
of enough bright NGSs for the WF sensing. This field is located at a
galactic latitude $b=-66.47\deg$, we have star positions and
magnitudes up to $R=19$. We chose $15$ random positions within this
field as the center of $5'$ diameter science sub-fields and NGS
pick-up fields. Fig.~\ref{15-subfields}a shows the average, minimum
and maximum EE for the 15 chosen sub-fields, assuming a NGS pick-up
FoV limited to $7.3'$, 5 NGS dedicated WFSs, a NGS limiting magnitude
of $R=17$. We observe a EE performance spread between $40$ and $75\%$
depending on the sub-field. We recall that considering only one bright
NGS in the science field, we found EE$=30\%$ for $6$ LGS and $67\%$
for $9$. In Fig.~\ref{15-subfields}a, even if the variability of EE is
important, it is possible most of the time to achieve EE greater or
close to $50\%$, slightly lower than the one with $9$ LGS and one
bright NGS, but a performance which could be sufficient depending on
the total WFE budget.  Fig.~\ref{15-subfields}b shows the sub-field
$1$ randomly selected in the XMM-LSS cosmologic field as an example
corresponding to relatively poor performance in
Fig.~\ref{15-subfields}a. In the sub-field of $7.3'$ in diameter, only
4 NGS brighter than $R=17$ are available with only one relatively
bright star $R=12.9$.

\newlength{\imagewidth}
\setlength{\imagewidth}{.42\linewidth}
\begin{figure}
  \begin{center}
    \begin{tabular}{cc}
      \includegraphics[height=\imagewidth]{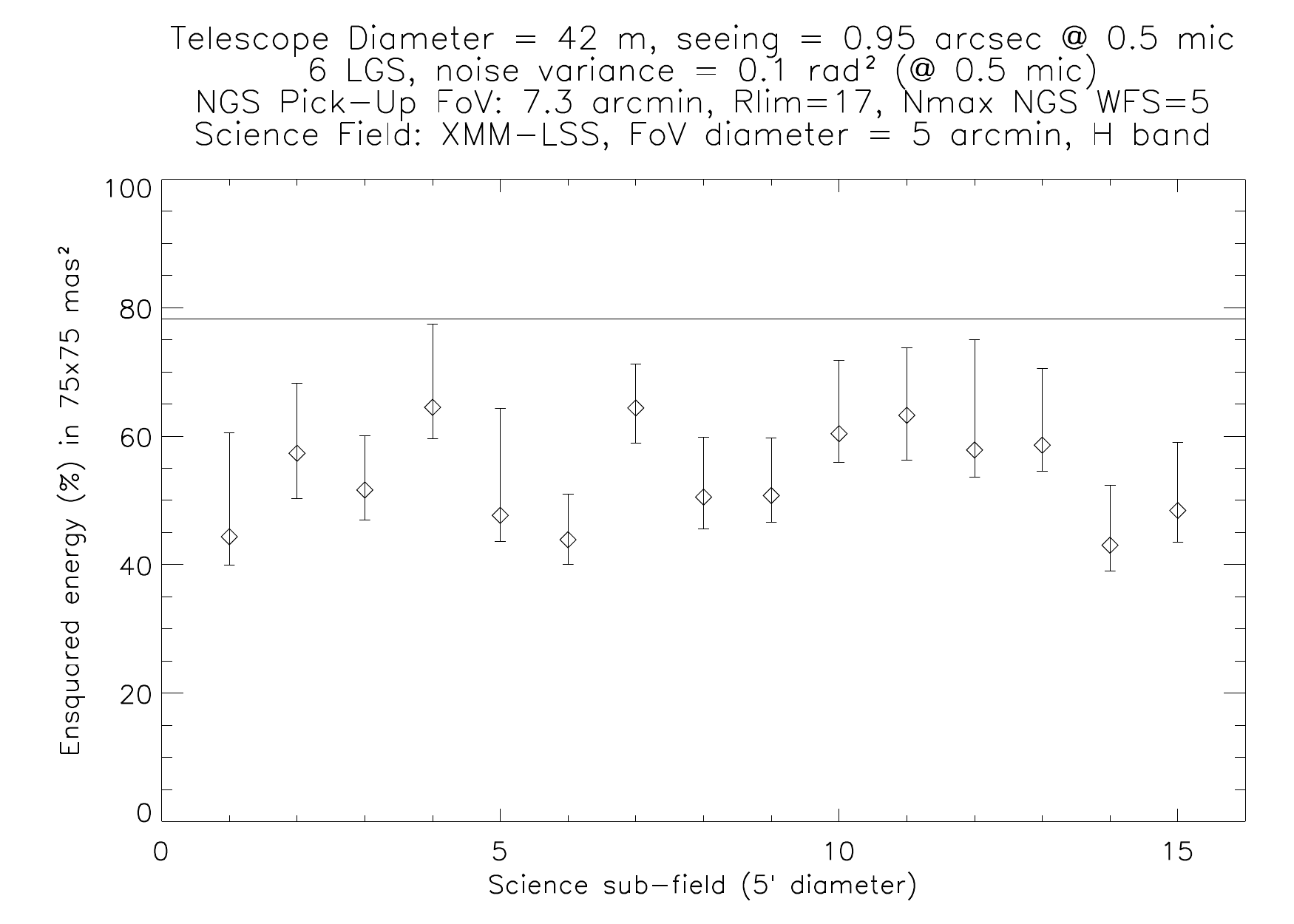} &
      \includegraphics[height=\imagewidth]{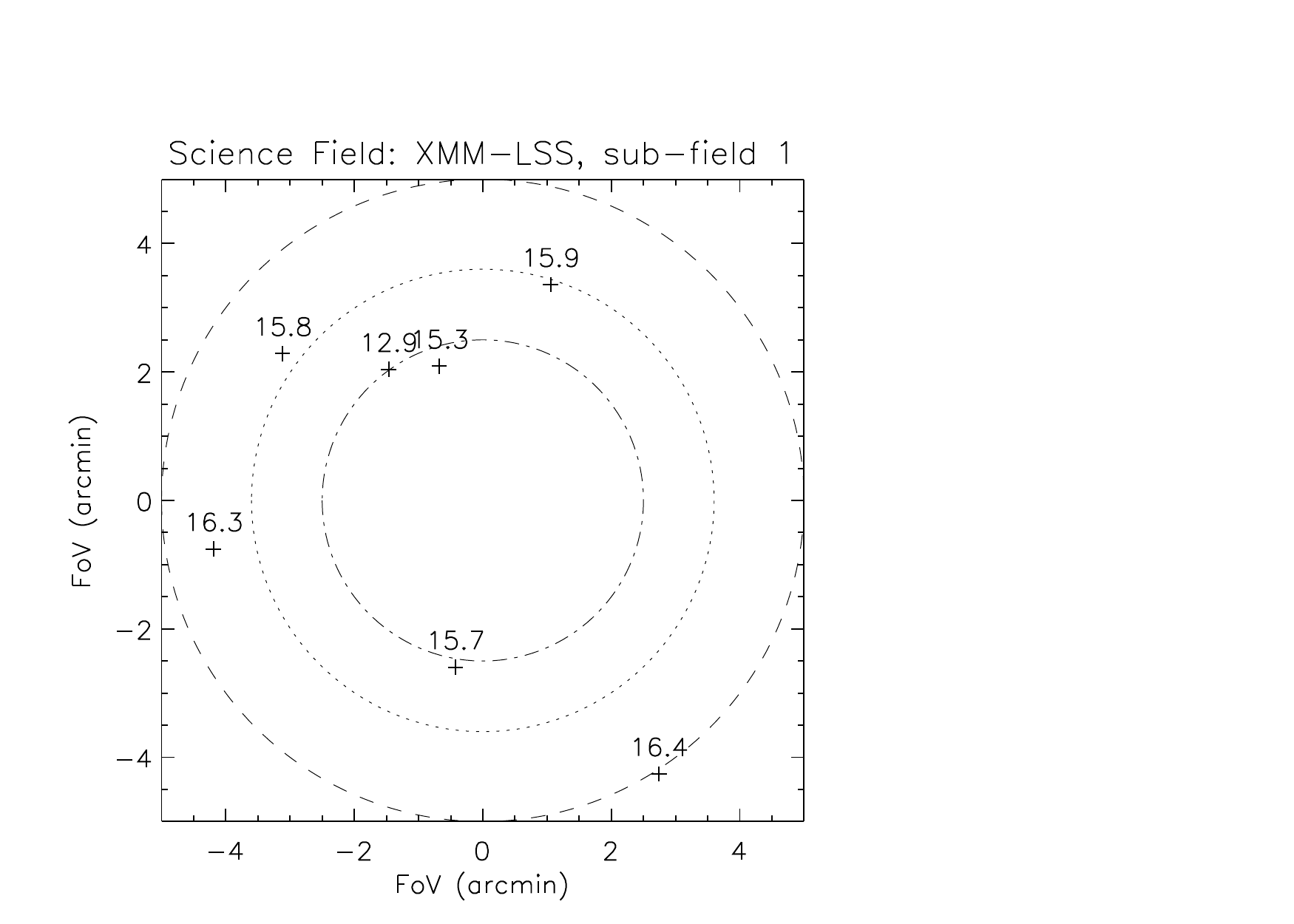} \\
      {[a]} & \hspace{-0.85in}{[b]}  \\
    \end{tabular}
    \caption{[a] Simulated EE as a function of the randomly selected science sub-fields in the XMM-LSS cosmologic field: diamond average value and spread bar between min. and max. values computed on the whole $5'$ sub-field. [b] One example of sub-fields. Crosses: NGS positions. Values: NGS magnitudes. Circle diameters: $10'$ for external, $7.3'$ for intermediate, $5'$ for internal.}
    \label{15-subfields}
  \end{center}
\end{figure}

In terms of sky coverage, we analysed the number of NGS available in a
given patrol field, for different regions of the sky using real target
fields. For fields randomly selected at galactic latitudes of $50 <
|b| < 90$, we obtain a chance larger than $80\%$ to find five or more
NGSs with $R<17$ in a $7'$ diameter field. There is a considerable
gain in NGS availability by exploiting patrol field beyond the central
$5'$ circular field. This result is also confirmed by star
statistics. Using the Besancon model, we obtain a probability, to find
five NGSs with $R<17$ in a $7.3'$ diameter field, larger than: $90\%$
at latitude $|b| = 60 \deg$ and $60\%$ at galactic pole. For EAGLE,
the strategy to take advantage of the large FoV to find around 5 NGSs
for WF sensing is a good one, leading to attractive sky coverage, even
at the galactic pole, and to the substantial reduction of the number
of LGSs ($9$ to $6$).

We decided also to consider median seeing conditions to finally define
the number of actuators and subapertures. We assume a seeing of
$0.87''$ at $0.5\mu m$ in the line of sight and an outer scale of
$25m$. We update the LGS measurement noise to be more realistic. In
the simulation, we consider now an equivalent uniform variance for the
WF measurement noise of $1 rad^2$ at $0.589\mu m$. It is deduced from
a fitting of the propagation through the tomographic reconstructor of
fratricide effects and spot elongations, inducing non uniform noise in
the pupil, for the case of a downscale telescope and the side
launching~\cite{Robert,Gratadour}. This variance corresponds to around
$500$ photons per subaperture and per frame. For a subaperture of
$50cm$ in size and at $500Hz$ frame rate, it typically corresponds to
a launched power of 10W per LGS. With these conditions, we analyse the
choice of the number of actuators and WFS subapertures. The number of
actuators on the science DMs is always equal to the number of
subapertures of the LGS WFSs and varies between $64$x$64$, $84$x$84$
and $112$x$112$. The number of NGS WFS subapertures is $32$x$32$ and
$64$x$64$. Figure~\ref{NumberActuators} shows the results of the
simulation in terms of EE for these parameters for the sub-field 1, a
worst case for the XMM-LSS field. It is found that in terms of
tomographic and noise errors there is no substantial gain to go to
$112$x$112$ actuators and even with $64$x$64$ the performance is quite
impressive EE$>50\%$. For this field, there is a marginal gain to
increase the number of subapertures for the NGS WFSs from $32$x$32$ to
$64$x$64$ because 3 NGSs on the 4 available are faint, magnitude
$R>15$. For more favorable cases, a typical $5\%$ addition in EE is
possible (for sub-field 10 for example). In conclusion, it is possible
most of time to obtain with 6 LGS and up to 5 NGSs in a field of
$7.3'$ diameter a EE larger than $>50\%$. We see that $84$x$84$ are
sufficient for the DMs and that $64$x$64$ could be a back-up solution
to the price of a slight decrease in performance.

\setlength{\imagewidth}{.35\linewidth}
\begin{figure}
  \begin{center}
  \begin{tabular}{cc}
  \includegraphics[height=\imagewidth]{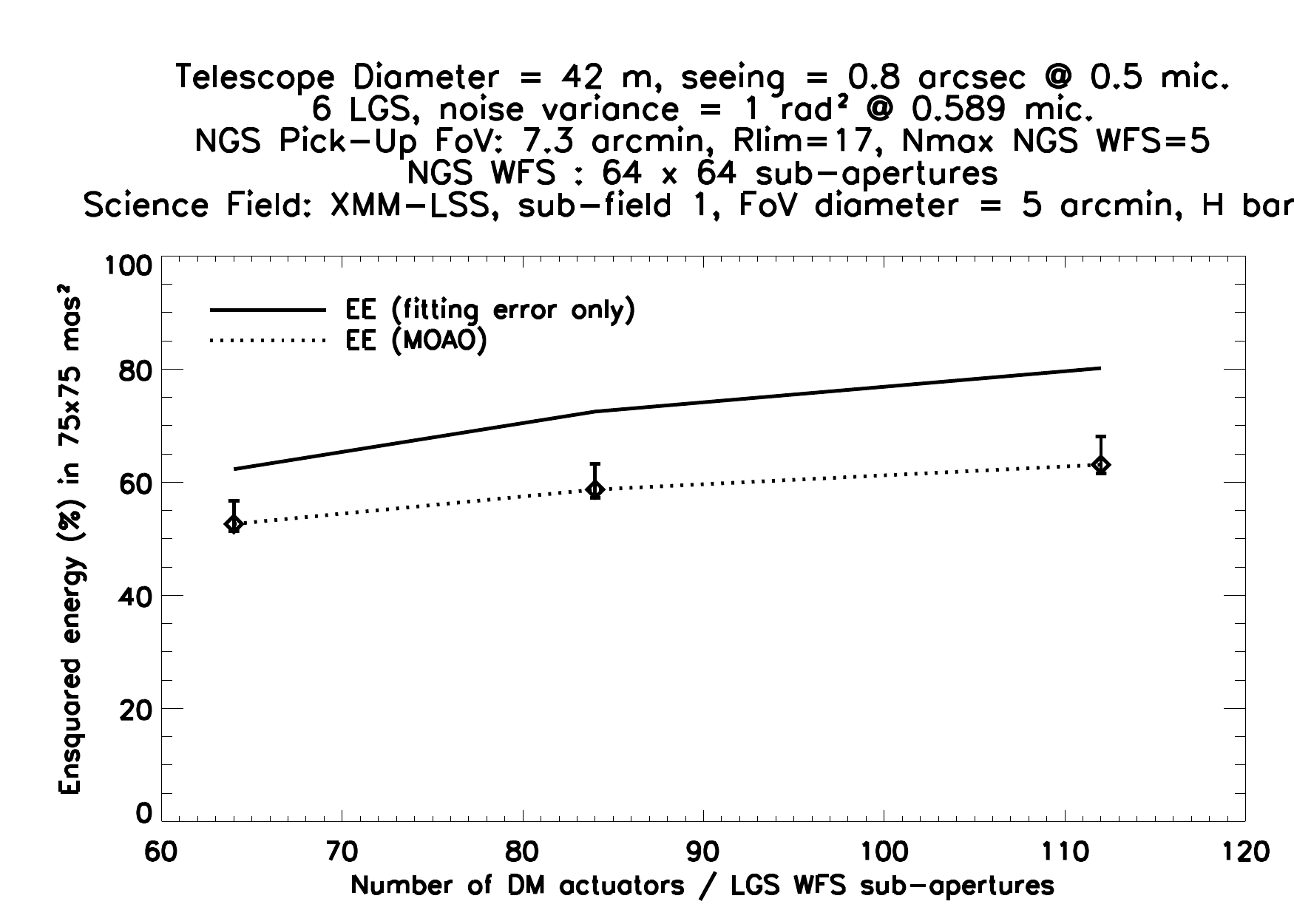} &
  \includegraphics[height=\imagewidth]{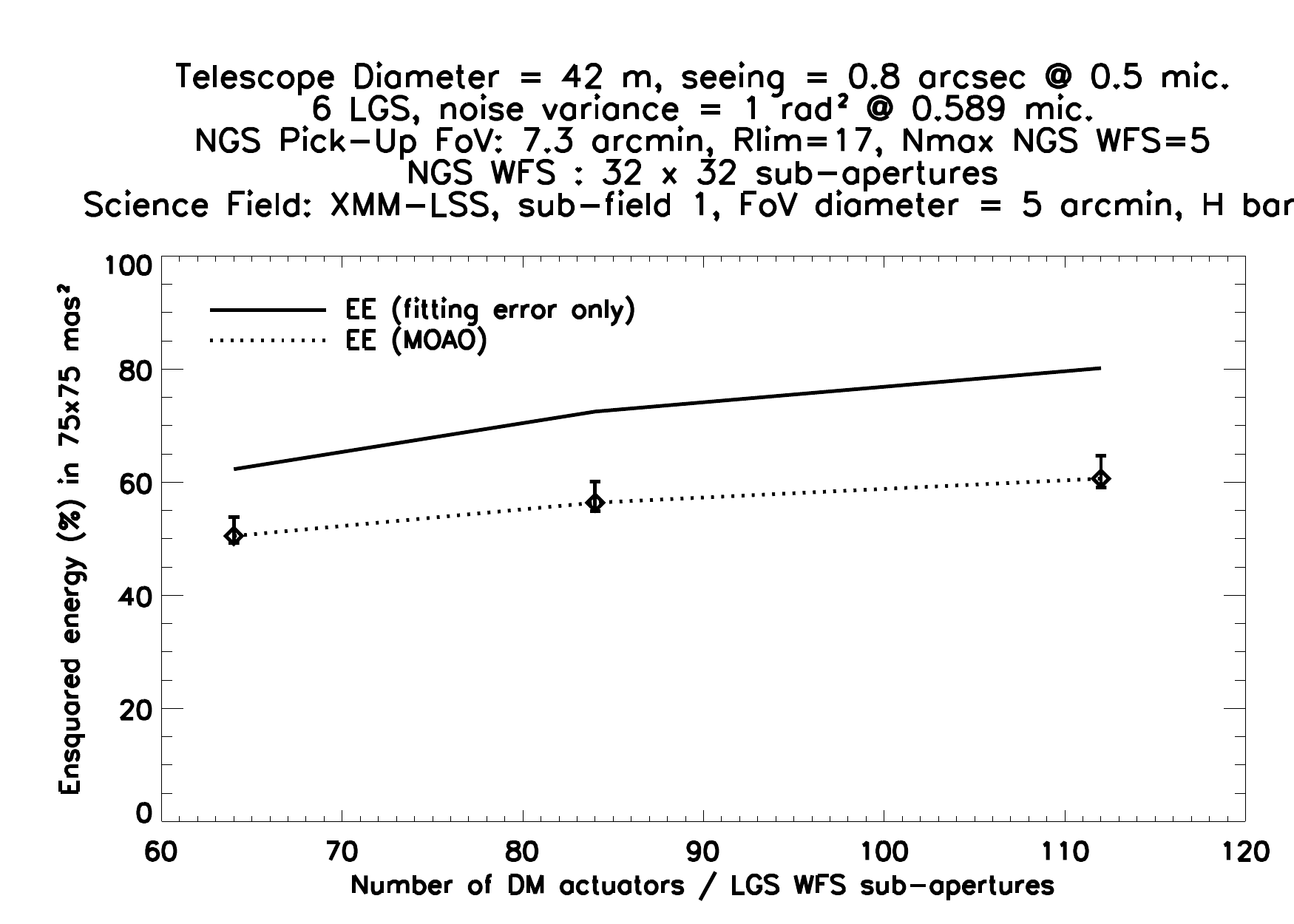} \\ 
  {[a]} & {[b]} \\
  \end{tabular} 
  \caption{Simulated EE as a function of the number of
  actuators (equal number of LGS WFS subapertures) for the sub-field 1
  in the XMM-LSS cosmologic field (worst case): [a] $64$x$64$ NGS WFS
  subapertures.  [b] $32$x$32$ NGS WFS subapertures. Diamond for
  average EE value and spread bar for min. and max. EE values,
  computed on the centred science $5'$ field. Continuous line: fitting
  error only} \label{NumberActuators} 
\end{center}
\end{figure}

\section{Preliminary AO system design}
\label{design}
The current design of the MOAO system for EAGLE is as followed.  To
ensure a good sky coverage and a relatively high performance, we
propose to implement 6 LGSs of 10W located on a $7.5'$ diameter ring
around the science FoV. The LGS WFS will be Shack-Hartmann with
$84$x$84$ subapertures. The LGS detector is quite a challenge, but
developments are conducted by ESO~\cite{Kolb}.  Up to 6 NGS WFSs of
medium order ($64$x$64$ as a maximum) will be available to ensure the
maximum covering of the $38.5arcmin^2$ patrol FoV. The NGSs will be
picked-up anywhere in this FoV. The magnitude of these stars should be
lower than $R=17$. The science channel DMs will be made of $84$x$84$
actuators. The existing $4k$ DM should be a back-up solution in case
of development problems for a larger count. The real time computer
will implement an optimal tomographic reconstructor and a dedicated
direction projector taking advantage of a $C^2_n$ profiling on site.
The sampling temporal frequency of the loop should be ajustable
between typically $250$ and $25Hz$. For calibrations, the main issue
in addition to convention needs is the requirement of a very good
pupil registration between all the parallel channels and its stability
during observations because of the open loop constraints. We propose
to use M4 (and some tools around to tag the pupil) to be able using
artificial sources inside the telescope to record simultaneously and
in a short time the current position of the image of M4 and the pupil
conjugated components of the system. It will require to install in the
science channels a WFS and a pupil imager to be used off-line only for
this purpose.

The error budget of the AO system is split in LO and HO contributors
having different impact on EE~\cite{Fusco}. Dedicated expressions
depending on the spatial frequencies allow us to link the WFE in $nm$
to the EE loss. In addition to the tomographic (including fitting and
aliasing) and noise error, we take into account the following main
errors: chromatism, refraction, differential focal anisoplanatism,
temporal bandwidth, turbulence model, open loop both on the WFSs and
on the DMs, calibrations and non common path aberrations, with some
additional contingencies. Starting from the $53\%$ EE in the worst
case given by the simulations presented in Section~\ref{analysis}, it
leads to an overall performance of $35\%$ EE well in the
specifications.

At system level, the MOAO is not yet tested on-sky, but there are
several on-going programs worldwide which will demonstrate its
principle of operations. Within the EAGLE framework, an aggressive
technology development plan is in place which will allow us to
demonstrate, test, and characterize MOAO within a couple of
years. Demonstration activities are taking place on a laboratory test
bench, SESAME, at Observatoire de Paris. Different DM technologies,
such as magnetic, piezoelectric or electrostatic actuators, have been
tested in terms of open loop behaviour. The laboratory measured open
loop errors for a few DMs are compatible with the requirement of
EAGLE: rms error of the order of $2$ to $4\%$ of the rms stroke used
for the compensation. The second key activity is the demonstration of
the open loop optimised tomographic compensation~\cite{Vidal}. The
laboratory sumulation is in the conditions of a 4m class telescope
with $3$ off-axis GSs, a piezoelectric DM in open loop in the on-axis
diagnostic channel and a seeing corresponding to $D/r_o=12.8$. For a
Strehl ratio of less than $1\%$ without correction, we obtain $39\%$
in a conventional closed loop scheme and $35\%$ in pure open loop MOAO
using an original tomography algorithm~\cite{Vidal}. These results are
the first stage improving the technology readiness. An on-sky
demonstration programme (CANARY) is being actively developed for use
at the WHT under the leadership of the Durham
University~\cite{Morris}, with the first results expected within 1
year with $3$ off-axis NGSs and one on-axis channel for performance
evaluation including $1$ piezoelectric DM in open loop, and within 2
years with $4$ additional Rayleigh LGSs, on time to validate the final
EAGLE design.

\section{Conclusion}
The baseline design of EAGLE, the multi-object spatially-resolved
near-IR spectrograph concept for the E-ELT, is driven by the
scientific requirements to answer a number of crucial questions about
how galaxies formed and evolved. MOAO is the only tractable AO concept
achieving the specifications on the $20$ parallel IFU subfields of
size $1.65''$x$1.65''$ in a very wide FoV. Using a $84$x$84$ actuator
deformable mirror for each target direction, our analysis demonstrates
that $6$ laser guide stars and up to $5$ natural guide stars of
magnitude $R<17$, picked-up in a $7.3'$ diameter patrol field of view,
are a good configuration for tomography. The computed performance is
EE$=35\%$ in a $75$x$75 mas^2$ spaxel at $H$ band whatever the target
direction in the centred $5'$ science field, for median seeing
conditions. The sky coverage at galactic latitudes $|b|\sim60$ is
close to $90\%$ from both real fields and star statistics.

\section{Acknowledgements}
We acknowledge support from CNRS/INSU (France) and STFC (UK) ; the
French Agence Nationale de la Recherche (ANR) program 06-BLAN-0191 ;
the European Southern Observatory (ESO) (phase A study of a Wide
Field, multi-IFU near IR Spectrograph and AO system for the E-ELT) ;
the European Commission (Framework Programme 7, E-ELT Preparation,
Infrastructure 2007-1 Grant 211257) ; the Observatoire de Paris.

\end{document}